\documentclass[epj]{svjour}
\usepackage{graphics}

\usepackage{cite}
\usepackage{epsf}
\usepackage{rotating}
\usepackage{here}

\def\ks{$K_S$}
\def\ksp{$K_S p$}
\def\la{$\Lambda$}
\def\lapip{$\Lambda \pi^+$}

\def\t1540{$\Theta^+ (1540)$}
\def\xf{$x_F$}

\begin{document}

\newcounter{str1}
\newcounter{str2}
\newcounter{str3}
\newcounter{str4}
\newcounter{str5}
\newcounter{str6}

\title{Observation of a resonance in the \ksp\ decay channel at a mass of 1765 MeV/c$^2$}

\titlerunning{Observation of a resonance in the \ksp\ decay channel}
\authorrunning{WA89 Collaboration}
\subtitle{WA89 Collaboration}
\author{
\renewcommand{\thefootnote}{{\rm\alph{footnote}}}%
\setcounter{footnote}{0}%
  M.I.~Adamovich              \inst{1,}\thanks{Deceased} \and
  Yu.A.~Alexandrov            \inst{1,}\thanks{Supported by Deutsche
  Forschungsgemeinschaft, contract number 436 RUS 113/465/0-2(R), and Russian
  Foundation for Basic Research under contract number RFFI 00-02-04018.}
  \setcounter{str3}{\value{footnote}} \and
  S.P.~Baranov                \inst{1} \fnmsep \footnotemark[\value{str3}] \and
  D.~Barberis                 \inst{2} \and
  M.~Beck                     \inst{3} \and
  C.~B\'erat                  \inst{4} \and
  W.~Beusch                   \inst{5} \and
  M.~Boss                     \inst{6} \and
  S.~Brons                    \inst{3,}\thanks{Now at TRIUMF, Vancouver, B.C., Canada V6T 2A3} \and
  W.~Br\"uckner               \inst{3} \and
  M.~Bu\'enerd                \inst{4} \and
  Ch.~Busch                   \inst{6} \and
  Ch.~B\"uscher               \inst{3} \and
  F.~Charignon                \inst{4} \and
  J.~Chauvin                  \inst{4} \and
  E.A.~Chudakov               \inst{6,}\thanks{Now at Thomas Jefferson Lab, Newport News, VA 23606, USA.} \setcounter{str4}{\value{footnote}}\and
  U.~Dersch                   \inst{3} \and
  F.~Dropmann                 \inst{3} \and
  J.~Engelfried               \inst{6,}\thanks{Now at Instituto de Fisica, Universidad Autonoma de
                                                San Luis Potosi, S.L.P. 78240 Mexico.} \and
  F.~Faller                   \inst{6,}\thanks{Now at Fraunhofer Institut f\"ur Solar Energiesysteme, D-79100 Freiburg, Germany.} \and
  A.~Fournier                 \inst{4} \and
  S.G.~Gerassimov             \inst{3,1,}\thanks{Now at Physik Department E18, Technische Universita\"at M\"unchen, D-85747 Garching, Germany.}
  \setcounter{str2}{\value{footnote}}\and
  M.~Godbersen                \inst{3} \and
  P.~Grafstr\"om              \inst{5} \and
  Th.~Haller                  \inst{3} \and
  M.~Heidrich                 \inst{3} \and
  E.~Hubbard                  \inst{3} \and
  R.B.~Hurst                  \inst{2} \and
  K.~K\"onigsmann             \inst{3,}\thanks{Now at Fakult\"at f\"ur Physik,
  Universit\"at Freiburg, Germany.}  \setcounter{str1}{\value{footnote}}  \and
  I.~Konorov                  \inst{3,1,} \footnotemark[\value{str2}] \and
  N.~Keller                   \inst{6} \and
  K.~Martens                  \inst{6,}\thanks{Now at Department of Physics, University of Utah, Salt Lake City, Utah, USA.} \and
  Ph.~Martin                  \inst{4} \and
  S.~Masciocchi               \inst{3,}\thanks{Now at Gesellschaft f\"ur Schwerionenforschung, D-64291 Darmstadt,
Germany.}\setcounter{str5}{\value{footnote}}\and
  R.~Michaels                 \inst{3} \fnmsep \footnotemark[\value{str4}]  \and
  U.~M\"uller                 \inst{7} \and
  H.~Neeb                     \inst{3} \and
  D.~Newbold                  \inst{8} \and
  C.~Newsom\thanks{University of Iowa, Iowa City, Iowa 52242, USA.} \and
  S.~Paul                     \inst{3} \fnmsep \footnotemark[\value{str2}] \and
  J.~Pochodzalla              \inst{3,7} \and
  I.~Potashnikova             \inst{3} \and
  B.~Povh                     \inst{3} \and
  Z.~Ren                      \inst{3} \and
  M.~Rey-Campagnolle          \inst{4,}\thanks{Permanent address: CERN, CH-1211 Gen\`eve 23, Switzerland.} \and
  G.~Rosner                   \inst{7,}\thanks{Now at Department of Physics and Astronomy, University of Glasgow, Glasgow G12 8QQ, United Kingdom.} \and
  L.~Rossi                    \inst{2} \and
  H.~Rudolph                  \inst{7} \and
  C.~Scheel\thanks{NIKHEF, 1009 D8 Amsterdam, The Netherlands.} \and
  L.~Schmitt                  \inst{7}\fnmsep \footnotemark[\value{str5}] \and
  H.-W.~Siebert               \inst{6,7} \and
  A.~Simon                    \inst{6}\fnmsep \footnotemark[\value{str1}] \and
  V.~Smith                    \inst{8} \and
  O.~Thilmann                 \inst{6} \and
  A.~Trombini                 \inst{3} \and
  E.~Vesin                    \inst{4} \and
  B.~Volkemer                 \inst{7} \and
  K.~Vorwalter                \inst{3} \and
  Th.~Walcher                 \inst{7} \and
  G.~W\"alder                 \inst{6} \and
  R.~Werding                  \inst{3} \and
  E.~Wittmann                 \inst{3} \and
  M.V.~Zavertyaev             \inst{1}\fnmsep \footnotemark[\value{str3}]
}

\mail{Josef Pochodzalla: pochodza@kph.uni-mainz.de}

\institute{
\renewcommand{\thefootnote}{{\rm\alph{footnote}}}%
    Moscow Lebedev Physics Institute, RU-117924, Moscow, Russia \and
    Dipartimento di Fisica and I.N.F.N, Sezione di Genova, I-16146 Genova, Italy \and
    Max-Planck-Institut f\"ur Kernphysik Heidelberg, D-69029 Heidelberg, Germany \and
    Institut des Sciences Nucl\'eaires, Universit\'e de Grenoble, F-38026 Grenoble Cedex, France \and
    CERN, CH-1211 Gen\`eve 23, Switzerland \and
    Physikalisches Institut, Universit\"at Heidelberg, D-69120 Heidelberg, Germany\thanks{Supported by the Bundesministerium f\"ur Bildung,
                      Wissenschaft, Forschung und Technologie,
                      Germany, under contract numbers 05~5HD15I, 06~HD524I
                  and 06~MZ5265.} \setcounter{str6}{\value{footnote}} \and
    Institut f\"ur Kernphysik, Universit\"at Mainz, D-55099 Mainz, Germany\footnotemark[\value{str6}] \and
    University of Bristol, Bristol BS8 1TL, United Kingdom
      }
\date{Received: 10.12.2000}
%
\abstract{ We report on the observation of a \ksp\  resonance signal
at a mass of  1765$\pm$5 MeV/c$^2$, with intrinsic width $\Gamma =
108\pm 22$ MeV/c$^2$, produced inclusively in $\Sigma^-$-nucleus
interactions at 340 GeV/c in the hyperon beam experiment WA89 at
CERN. The signal was observed in the kinematic region $x_F>0.7$, in
this region its production cross section rises approximately
linearly with $(1-x_F)$, reaching $BR(X\rightarrow K_S p)\cdot
d\sigma /dx_F\, =\,(5.2\pm 2.3)\, \mu b $ per nucleon at $x_F=0.8$.
The hard \xf\ spectrum suggests the presence of a strong leading
particle effect in the production and hence the identification as a
$\Sigma^{*+}$ state. No corresponding peaks were observed in the
$K^- p$ and $\Lambda \pi^{\pm}$ mass spectra.
} 
\maketitle
%

\setcounter{page}{1}
\section{Introduction}

Baryon spectroscopy has seen great progress in the charmed baryon
sector during the last years, but almost nothing has happened in the
non-strange and strange baryon sectors since about 1990 (see, for
instance, the review articles in the Review of Particle Physics
\cite{pdg}). In the strange baryon sector, the review lists 13
$\Lambda^*$, 9 $\Sigma^*$ and 5 $\Xi^*$ resonances classified as
``established'' (3 or 4 stars).

In particular in the $\Sigma^*$ sector, almost all known states
above the $\Sigma(1385)$ have been observed through partial wave
analyses only, with often widely varying estimates of the masses and
widths of the observed resonances. So far, there has been only one
observation of a $\Sigma^*$ above the $\Sigma(1385)$  in an
invariant mass spectrum: we have observed a wide peak at 1660
MeV/c$^2$ in the mass spectra of $\Lambda\pi^-$ pairs and, less
prominently, $\Lambda\pi^+$ pairs. These pairs were produced
inclusively in a copper/carbon target by a $\Sigma^-$ beam of 340
GeV/c momentum \cite{oursigmastar}.

We have also undertaken a high-statistics search for the pentaquark
candidate \t1540\ in the \ksp\ decay channel \cite{wa89theta}. While
no resonance signal was seen in the \ksp\ mass spectrum around 1540
MeV/c$^2$, we observed a \ksp\ resonance signal at 1765 MeV/c$^2$,
which is the topic of this publication.

\section{Event Selection}

The hyperon beam experiment WA89 at CERN ran from 1990 to 1994 in
the West Hall. The results presented here are based on the data
collected in the years 1993 and 1994. Details on the hyperon beam
and the experimental apparatus can be found in
\cite{wa89beam,beusch,wa89rich}.

The events were selected as described in detail in our previous
search for the \t1540\ pentaquark \cite{wa89theta}. \ks\ were
reconstructed in the decay $K^0_S \rightarrow \pi^+\pi^- $, using
all pairs of positive and negative particles which formed a decay
vertex in the decay zone. The contamination caused by $\Lambda
\rightarrow p\pi^-$ decays was reduced by excluding candidates with
a reconstructed $p\pi^-$ mass within $\pm 2 \, \sigma_m(\Lambda )$
of the \la\ mass ($\sigma_m(\Lambda )$ was 1.8 MeV/c$^2$ at low
momenta and 2.8 MeV/c$^2$ at 200 GeV/c). This requirement reduced
the \ks\ signal by 3\% and the background by 1/3 to less than {1\%}.

\begin{figure}[ht]
\centering
\includegraphics[width=0.9\linewidth]{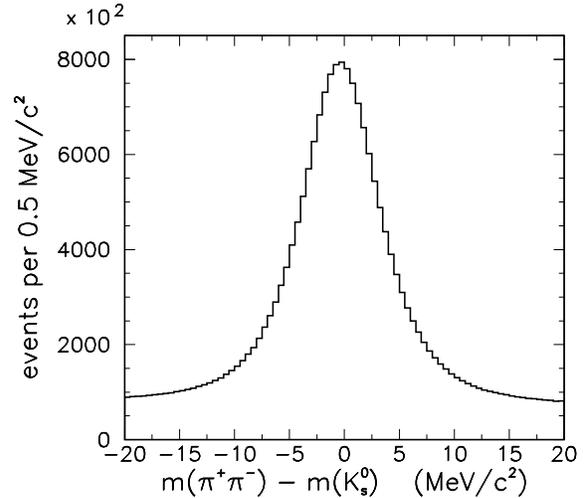}
\caption{Reconstructed $\pi^+ \pi^-$ mass distribution of the \ks\
candidates} \label{fig:ks}
\end{figure}

The reconstructed  $\pi^+ \pi^-$ mass distribution of the remaining
\ks\ candidates is shown in fig. \ref{fig:ks}. The peak from \ks\
decays contains about 13 million events, their momentum spectrum
extends from 10 GeV/c to about 200 GeV/c. Candidates with  a
reconstructed $\pi^+\pi^-$ mass within $\pm 2\, \sigma_m$ of the
\ks\ mass were retained for further analysis.

Proton candidates were selected from all additional positive
particles with a reconstructed track extending from the microstrip
counters downstream of the target to the wire chambers beyond the
spectrometer. Requiring  track reconstruction in the microstrip
counters rejected most of the protons from \la\ decays. The RICH
counter was used to purify the proton candidate sample. We required
the proton momentum to be  $p_p >45$ GeV/c, well above the proton
threshold of the RICH at 38 GeV/c \cite{wa89theta}.

\section{Results}

\begin{figure}[ht]
\centering
\includegraphics[width=0.9\linewidth]{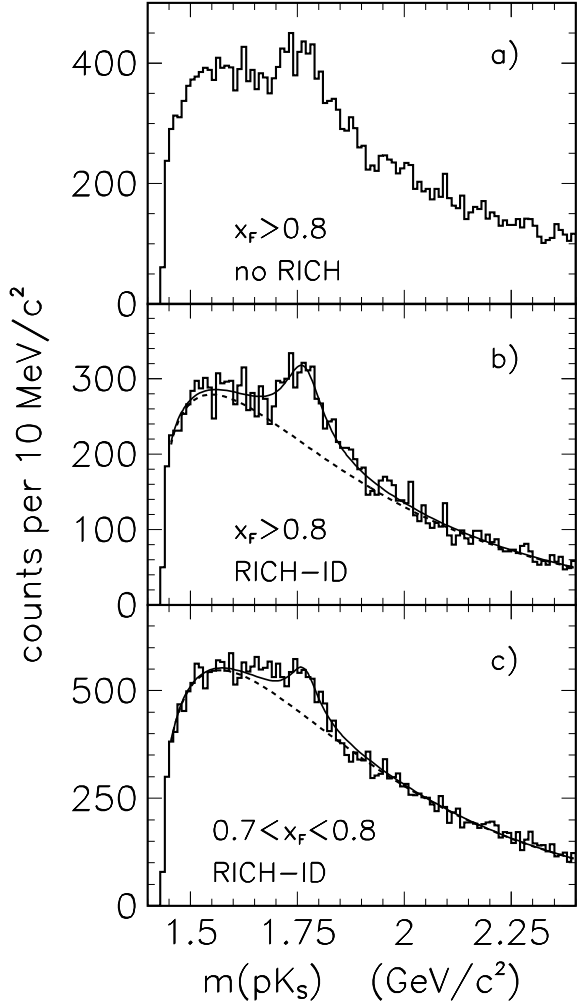}
\caption{\ksp\ invariant masses at high \xf . Proton identification
was required for b) and c).} \label{fig:kspr}
\end{figure}

Even without using proton identification by the RICH,
we noted a clear \ksp\ mass peak at large values of Feynman--x.
Fig. \ref{fig:kspr}a shows the \ksp\ mass spectrum for
$x_F>0.8$. In  Fig. \ref{fig:kspr}b we show the mass spectrum after
application of the RICH-cut for proton identification (see above).
 The peak again is clearly visible,
with approximately equal statistics, while the background is reduced
by 25\%. This proves the identification as a \ksp\ resonance signal.
The signal still appears as a less significant shoulder in the
region  $0.7<x_F<0.8$, with comparable statistics, as shown in  fig.
\ref{fig:kspr}c. At $x_F<0.7$, the signal disappears in the rapidly
increasing background.

The distributions were fitted by the sum of a S-wave Breit-Wigner
function for the signal and a function $f(m)=C\cdot  (m-m_0)^a \cdot
e^{-(m-m_0)/b}$ for the background, where $m_0$ is the \ksp\
threshold and C, a and b are fit parameters. The solid lines show
the fit result for signal plus background
 and the dashed lines
show the background. In the region $x_F>0.8$, the resonance contains
N = 1380 $\pm$ 260 events, where the error includes the errors from
the choice of background parametrization and fit window. The mass of
the resonance is $m\, =\,  1765 \pm 5$ MeV/c$^2$ and its width is
$\Gamma\, =\, 108\pm 22$   MeV/c$^2$. The fit has $\chi^2 /NDF\, =\,
106/89$. Fitting the distribution to the background function $f(m)$
only, we obtained    $\chi^2 /NDF\, =\, 226/92$. For $0.7<x_F<0.8$,
we obtained N = 1560 $\pm$ 340 events and values of m and $\Gamma$
compatible with those for $x_F>0.8$. Replacing $f(m)$ in the fit
with the observed shape of the \ksp\ mass distribution of ``mixed''
events, leaving only the background normalization as a fit
parameter, we found, for $x_F>0.8$, an increase of the number of
events by 8\%, well within the errors from the fit using $f(m)$.

The production cross section per nucleon was calculated assuming its
dependence on the mass number A of the target nucleus to be
$\sigma(A) \propto A^{2/3}$. The result is $BR(X\rightarrow K_S
p)\cdot d\sigma /dx_F\, =\, (5.2\pm 2.3)\, \mu b$ {\it per nucleon},
at \xf =0.8. The error on the cross section includes the
uncertainties of the trigger efficiency for this decay mode. The
$x_F$ dependence can be parametrized as $d\sigma  /dx_F \propto
(1-x_F )^n$, with $n\, =\, 1.0\pm 0.5$.

\begin{figure}[ht]
\centering
\includegraphics[width=0.9\linewidth]{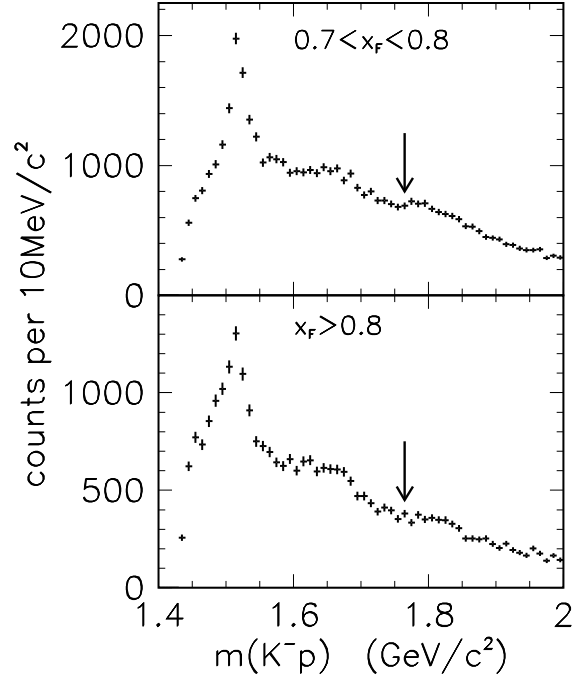}
\caption{$K^- p$ invariant masses at  high \xf . The arrows indicate
the position of the \ksp\ peak.} \label{fig:kmpr}
\end{figure}

We have searched for an isospin partner of this resonance in the
$K^-p$ decay channel. Candidates for this decay had to have
 a $K^-$ and a proton candidate with their resp. tracks
emerging from the target and with RICH identification. For the
proton candidates we used the same momentum and RICH cuts as those
used for the \ksp\ sample. For the $K^-$ we used analogous cuts,
$p_K>25$ GeV/c and a RICH-cut with the same efficiency. The
resulting $K^-p$ mass spectra are shown in fig. \ref{fig:kmpr}. They
are dominated by the $\Lambda(1520)$ peak. Less significant
structures are visible around 1650 and 1800 MeV. These structures
have also been observed by the SPHINX collaboration \cite{sphinx}
and can be attributed to known $\Lambda^*$ or $\Sigma^*$ resonances
\cite{pdg}. No evidence for a resonance around 1765 MeV/c is
visible, also not at lower $x_F$. Upper limits for the cross section
were calculated treating the mass and width of the resonance with
their experimental errors as nuisance parameters (see e.g.
\cite{Feldman}). We obtained a limit of\\
\centerline{$\sigma\cdot BR(K^-p)/ \sigma\cdot BR( K_Sp) \, < 0.6 \,
(95 \% \, CL)$.}

\begin{figure}[ht]
\centering
\includegraphics[width=\linewidth]{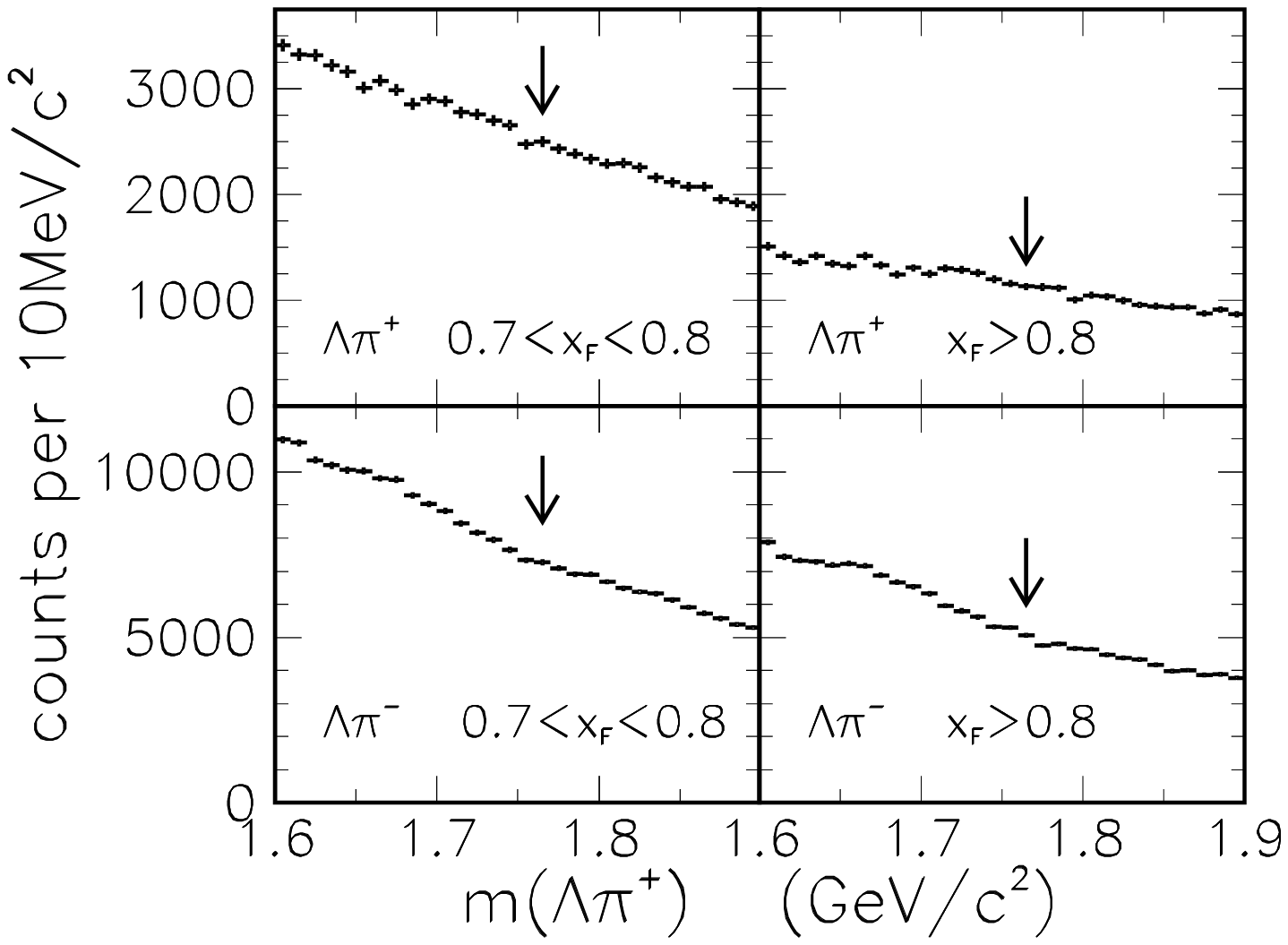}
\caption{$\Lambda \pi$ invariant masses at  high \xf . The arrows
indicate the position of the \ksp\ peak.} \label{fig:lapi}
\end{figure}

We have also searched for this resonance and a possible isospin
partner in the $\Lambda\pi^{\pm}$ decay channels. Candidates for
$\Lambda\rightarrow p\pi^-$ decays were selected with criteria
analogous to those used in the selection of $K_S\rightarrow
\pi^+\pi^-$ candidates. All additional charged particles emerging
from the target with a reconstructed track in the microstrip
counters were considered as pion candidates. Since this sample
contained a large fraction of low-momentum pions, a cut $\cos
\vartheta^*(\pi)>-0.8$ was applied, which rejected about 50\% of the
$\Lambda\pi^+$ and 25\% of the $\Lambda\pi^-$ candidates at
$x_F>0.7$ and $1700<m(\Lambda\pi)<1820$ MeV/c$^2$. The resulting
samples are shown in fig. \ref{fig:lapi}. Again, no  evidence for a
resonance around 1765 MeV/c$^2$ is visible, also not at lower \xf .
The following upper limits were obtained:
\centerline{$BR(\Lambda\pi^+ )/BR(K_S p) \, < 1.8  \, (95 \% \,
CL)$,} \centerline{$\sigma\cdot BR(\Lambda\pi^- )/ \sigma\cdot BR(
K_Sp) \, < 3.4  \, (95 \% \, CL)$.}

\section{Discussion}
The strangeness of the observed resonance could be S=+1 or S=--1.
In the first case, this would be an exotic state like the pentaquark candidate
\t1540 . The assignment S=--1 leads to the more likely interpretation
as a  $\Sigma^{*+}$, which we will discuss further.

In table \ref{sigmaprod} we list the differential cross sections
{\it per nucleon} at $x_F=0.75$ for production of $\Sigma^{\pm}$ and
$\Sigma^{*\pm }$ measured in our experiment \cite{oursigmastar}. The
$x_F$-dependence of the  $\Sigma^+(1765)$ production cross section
is very similar to  that of the $\Sigma^+$ and  $\Sigma^+(1385)$
production. This fact suggests that also in  $\Sigma^+(1765)$
production at high \xf , a strange quark is transferred from the
beam projectile to the outgoing hyperon, which supports the S=--1
assignment to this state. (For a detailed discussion of the
``leading particle effect'' in hadronic hyperon production, see ref.
\cite{oursigmastar}).

\begin{table}[h]
\centering
\begin{tabular}{|c|c|c|c|}
\hline
& $d\sigma /dx_F$ & $BR\cdot d\sigma /dx_F$ & \\
& \multicolumn{2}{c|}{[$\mu b$]} & n  \\  \hline \hline
$\Sigma^+$ & 850$\pm$200 & & 1.7$\pm$0.2   \\
$\Sigma^-$ & 15500$\pm$1500 & & -0.2$\pm$0.2   \\  \hline
$\Sigma^+ (1385)$ & 600$\pm$100 & & 1.0$\pm$0.2   \\
$\Sigma^- (1385)$ & 1500$\pm$200 & & 0.5$\pm$0.1   \\  \hline
$\Sigma^+ (1660) \rightarrow \Lambda \pi^+$ & & 100$\pm$50& 4$\pm$1 \\
$\Sigma^- (1660) \rightarrow \Lambda \pi^-$ & & 550$\pm$100& 0.5$\pm$0.1\\
\hline
$\Sigma^+ (1765) \rightarrow K_S p$ & &  6$\pm$3 & 1.0$\pm$0.5  \\
$\Sigma^0 (1765) \rightarrow K^- p$ & &  $<3.5$ &  \\
$\Sigma^+ (1765) \rightarrow \Lambda\pi^+$ & & $<11 $  &   \\
$\Sigma^- (1765) \rightarrow \Lambda\pi^-$ & & $<20 $  &   \\
   \hline
\end{tabular}
\caption{$\Sigma$ and $\Sigma^*$ production cross sections at $x_F
=0.75$, measured in WA89. $d\sigma /dx_F$ is per nucleon and is
parametrized as $d\sigma / dx_F \propto (1-x_F) ^n$. }
\label{sigmaprod}
\end{table}

Numerous candidates for $\Sigma^*$ states have been found in PWA
analyses \cite{pdg}. There are two known states in the region of
1765 MeV/c$^2$, which are shown in table \ref{sigmastar}. The
parameters of our observed signal are compatible with either state,
so it could well be one of both or a combination of them. Our limit
on the \lapip\ decay mode is too high to enable a separation of the
two states.

\begin{table}[h]
\centering
\begin{tabular}{|c|c|c|}
\hline
 & $\Sigma(1750)\, S_{11} $ & $\Sigma (1775)\, D_{15} $ \\ \hline
 status & *** & ****\\
 mass (MeV/c$^2$) & 1760$\pm$10 & 1775$\pm$10 \\
width $\Gamma$  (MeV/c$^2$)& 60 - 160 & 135$\pm$20 \\
BR(NK) & 0.1 - 0.4 & 0.40$\pm$0.03\\
BR($\Lambda\pi$) & seen & 0.17$\pm$0.03 \\ \hline
\end{tabular}
\caption{The basic data of the $\Sigma(1750)$ and $\Sigma(1775)$
from the Particle Data Group \cite{pdg}.} \label{sigmastar}
\end{table}

So far, no $\Sigma^*$ resonance above the $\Sigma(1385)$ has been
observed directly in a mass plot with the exception of the
$\Lambda\pi^{\pm}$ resonances observed at lower \xf\ in our
experiment at around 1660 MeV/c$^2$ \cite{oursigmastar}. This
$\Lambda \pi^-$ resonance is visible even at $x_F>0.8$, as shown on
the bottom right of fig. \ref{fig:lapi}.

From the data on $\Sigma^{\pm}(1385)$ and  $\Sigma^{\pm}(1660)$
production, which exhibit a strong leading particle effect
\cite{oursigmastar}, we would expect the production cross section
for $\Sigma^-(1765)$ to be three to five times larger than for
$\Sigma^+(1765)$. $\Sigma^0(1765)$ production then would probably
also be enhanced w.r.t. $\Sigma^+(1765)$ production. In the relevant
kinematic range the acceptance for $\Sigma(1765)\rightarrow
\Lambda\pi$ is about a factor 0.65 lower than for
$\Sigma(1765)\rightarrow K_Sp$. This difference is mainly due to the
different decay lengths of $\Lambda$ and $K_S$ and to the fact that
the $\Lambda$ momenta resulting from the $\Lambda\pi$ decays are
larger than the $K_S$ momenta from the \ksp\ decays. Thus the
missing peak in the $\Lambda\pi$ channel can not be attributed to a
very different acceptance. However,  since the branching ratios to
the various decay modes of the $\Sigma^*$ isospin triplet are poorly
known, our limits on the $K^- p$ and $\Lambda\pi^{\pm}$ decays
(Tab.~\ref{sigmaprod}) cannot be translated into cross section
limits significant enough for a statement on the leading particle
effect.

To summarize: We have observed a \ksp\ resonance signal at
1765$\pm$5 MeV/c$^2$, with an intrinsic width $\Gamma = 108{\pm}22$
MeV/c$^2$. The \xf -dependence of the production cross section
favors the assignment of S=--1 to the resonance. Therefore the most
likely interpretation is that it is one or both of two $\Sigma^*$
resonances of similar mass and width, which are known from PWA. We
have not found this resonance or its possible  isospin partners in
the $K^- p$ and $\Lambda\pi^{\pm}$ decay channels, but the poor
knowledge of the branching ratios involved prevents any further
conclusions about the nature of this resonance.

\section*{ Acknowledgements }

We are indebted to J.~Zimmer and the late Z.~Kenesei for their help
during all moments of detector construction and setup. We are
grateful to the staff of CERN's EBS group for providing an excellent
hyperon beam channel, to the staff of CERN's Omega group for their
help in running the $\Omega $-spectrometer. We thank S.U. Chung for
valuable discussions.

\end{document}